**Gate tunable quantum Hall effects in defect-suppressed $Bi_2Se_3$ films**

Nikesh Koirala[1†*], Maryam Salehi[2], Jisoo Moon[1], Seongshik Oh[1*]

[1]Department of Physics & Astronomy, Rutgers, The State University of New Jersey, Piscataway, New Jersey 08854, U.S.A.

[2]Department of Materials Science and Engineering, Rutgers, The State University of New Jersey, Piscataway, New Jersey 08854, U.S.A.

[*]Correspondence should be addressed to nikesh@mit.edu and ohsean@physics.rutgers.edu

[†]Present Address: Department of Physics, Massachusetts Institute of Technology, Cambridge, MA 02139 U.S.A.

**Despite many years of efforts, attempts to reach the quantum regime of topological surface states (TSS) on an electrically tunable topological insulator (TI) platform have so far failed on binary TI compounds such as $Bi_2Se_3$ due to high density of interfacial defects. Here, utilizing an optimal buffer layer on a gatable substrate, we demonstrate the first electrically tunable quantum Hall effects (QHE) on TSS of $Bi_2Se_3$. On the *n*-side, well-defined QHE shows up, but it diminishes near the charge neutrality point (CNP) and completely disappears on the *p*-side. Furthermore, around the CNP the system transitions from a metallic to a highly resistive state as the magnetic field is increased, whose temperature dependence indicates presence of an insulating ground state at high magnetic fields.**

TSS provides a rich playground for a number of topological quantum effects such as topological magnetoelectric effect, Majorana fermions and QHE[1-8]. However, due to high level of surface Fermi level originating from unintended native dopants, it has been challenging to access the quantum regime of TSS. In particular, even if QHE is one of the most intensely studied phenomena in 2D systems, gate tuned studies of QHE of TIs has been so far limited only to ternary or quaternary compounds such as $Bi_{2-x}Sb_xTe_3$ and $(Bi_xSb_{2-x}Se_yTe_{3-y})$[4,5,9]. On the other hand, binary compounds can potentially provide cleaner platforms compared with quaternary or tertiary solid solutions, allowing better access to the Dirac point, due to increased carrier mobilities and reduced electron-hole puddles. For the binary compounds such as the prototypical TI $Bi_2Se_3$, tracking evolution of QHE as a function of gate voltage has not been possible due to high density of bulk and interfacial defects[10,11]. Recently, however, utilizing a structurally and chemically-matched buffer layer that solves the defect problem, QHE was observed in $Bi_2Se_3$ thin films[12,13]. Here, by adapting this buffer layer scheme to a gatable $SrTiO_3(111)$ substrate[14], we present the first gate-dependent study of QHE in $Bi_2Se_3$.

Low-carrier density $Bi_2Se_3$ thin films were grown on an electrically insulating buffer layer, which comprises a heterostructure of 5 QL $In_2Se_3$ – 4 QL $(Bi_{0.5}In_{0.5})_2Se_3$ grown on $SrTiO_3$ (111) substrate following the recipe of ref. 9 [15]. The films were then capped *in situ* either by a 100 nm Se or a 50 nm $MoO_3$/50 nm Se layer to protect against ambient contamination[16]. Then, a ~100 nm thick Cu layer was deposited *ex situ* on the back surface of $SrTiO_3$ substrate as a back gate. The films were then scratched into millimeter sized Hall bars using a metal mask and a tweezer, and indium leads were used to make electrical contacts [15].

On these Hall bar patterns, Hall ($R_H$) and sheet resistance ($R_S$) were initially measured at magnetic field ($B$) up to ± 0.6 T in a cryostat at $T$ = 5 K. The measured raw data were symmetrized or anti-symmetrized to eliminate mixing of longitudinal and Hall resistances due to imperfection in the measurement geometry [15]. $[e \cdot (dR_H/dB)]^{-1}$, which corresponds to sheet carrier density ($n_S$) for single-carrier species transport, and mobility ($\mu$) = $(R_o \cdot n_S \cdot e)^{-1}$ were then calculated, where $dR_H/dB$ is the slope of low field linear part of Hall resistance (unless otherwise stated), $e$ is the electronic charge and $R_o$ is the zero-field sheet resistance. These films have $n$-type carriers with $n_S \approx 5 \times 10^{12}$ cm$^{-2}$ and $\mu \approx 1{,}000 - 3{,}000$ cm$^2$V$^{-1}$s$^{-1}$ [15]. Compared to the films grown directly on SrTiO$_3$(111), where $n_S$ is typically ~$4 \times 10^{13}$ cm$^{-2}$, the buffer-layer grown films exhibit an order of magnitude decrease in the defect density[17], which is consistent with our previous report[12]. This low sheet carrier density obtained with the buffer layer was essential for reaching the quantum regime of TSS via gating as we present below.

In the rest of the paper, we focus on the gate voltage dependence of $R_H$ and $R_S$ in MoO$_3$/Se-capped films. As reported previously, MoO$_3$ capping further reduces the $n$-type Fermi level of Bi$_2$Se$_3$ films toward the charge neutrality point (CNP)[12,18]. We measured films of three different thicknesses: 8, 10 and 15 QL. 8 QL film was measured at $T = 1.5$ K and $B$ up to ± 9 T and 10 QL and 15 QL films were measured at $T = 5$ K and $B$ up to ± 0.6 T.

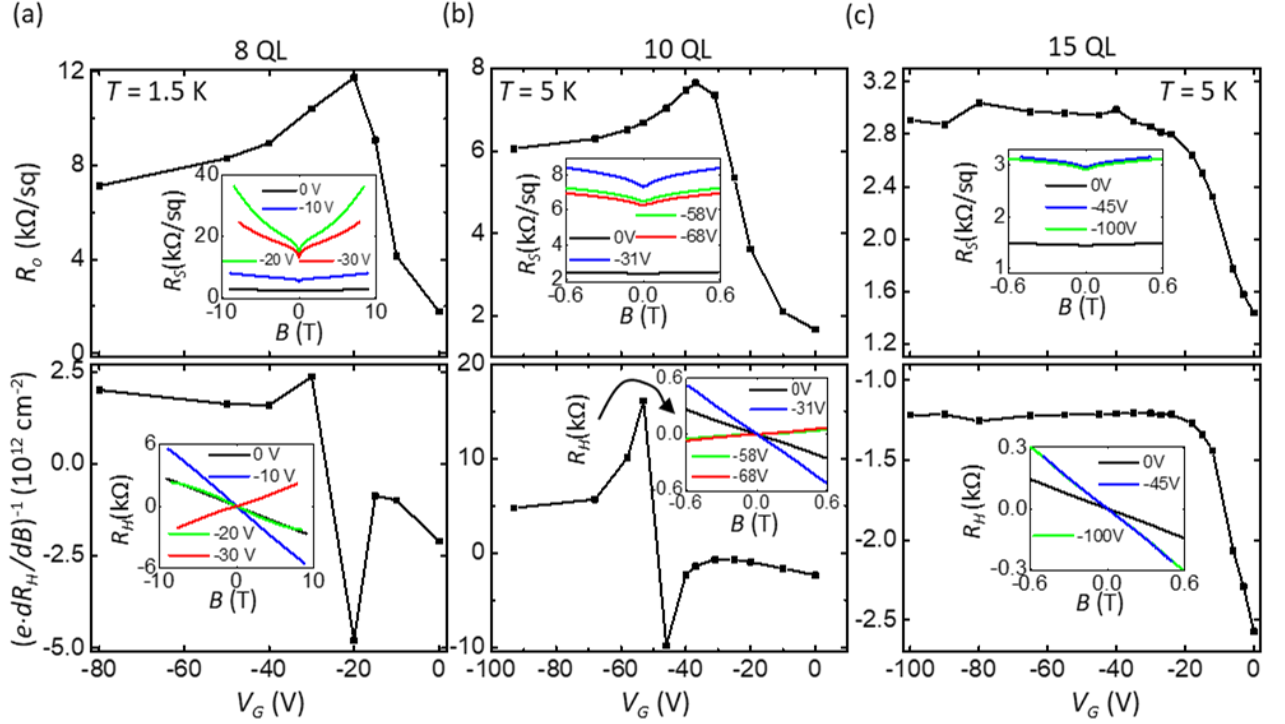

**Figure 1 | Gate-dependent magneto-transport data for $Bi_2Se_3$ films. (a)-(c)**, Zero-field sheet resistance ($R_o$, upper panel) and $(e \cdot dR_H/dB)^{-1}$ (lower panel), which corresponds to sheet carrier density ($n_S$) for single carrier species transport as a function of back-gate voltage, $V_G$, for 8, 10 and 15 QL films respectively. Solid black lines are a guide to the eye. The insets show magnetoresistance ($R_S$, upper panel) and corresponding Hall resistance ($R_H$, lower panel) as a function of magnetic field, $B$, taken at several representative back-gate voltage values from which $R_o$ and $n_S$ were extracted. Note the different magnetic field range and temperature for 8 QL compared to 10 and 15 QL films. Ambipolar behavior is observed in (a-b).

Figures 1 (a)-(c) show $R_o$ (upper panel) and $(e \cdot dR_H/dB)^{-1}$ (lower panel; for single species transport it is equivalent to $n_S$) as a function of back-gate voltage ($V_G$) of 8, 10 and 15 QL films, respectively. In all three samples, the *n*-type carrier density is less than $3 \times 10^{12}$ cm$^{-2}$ at $V_G = 0$ V, which is well below the maximum carrier density ($\sim 1 \times 10^{13}$ cm$^{-2}$) required to make the bulk state of TIs insulating[19]. For the 15 QL film, we were able to modulate $n_S$ from $2.8 \times 10^{12}$ cm$^{-2}$ to $1.5 \times 10^{12}$ cm$^{-2}$ and $R_o$ from ~1.5 kΩ/sq to ~2.8 kΩ/sq, as $V_G$ is tuned from 0 to -20 V. However, ambipolar transport was not observed in this film, presumably because 15 QL is too thick for its top surface to be electrostatically modulated by the bottom gating. For 8 and 10 QL films, $R_o$ increases with $V_G$, reaches a maximum value (for example at $V_G \approx -37$ V for 10 QL film) and then decreases with further increase in $V_G$. Concurrently, ($n$-type) $n_S$ decreases with $V_G$ and eventually changes to $p$-type (for example at $V_G \approx -45$ V for 10 QL film). Such an ambipolar behavior not only confirms the TSS conduction, but also the tunability of chemical potential across the CNP[20]. Therefore, we focus on 10 QL films below and carry out more in-depth measurements up to much higher magnetic fields (34.5 T).

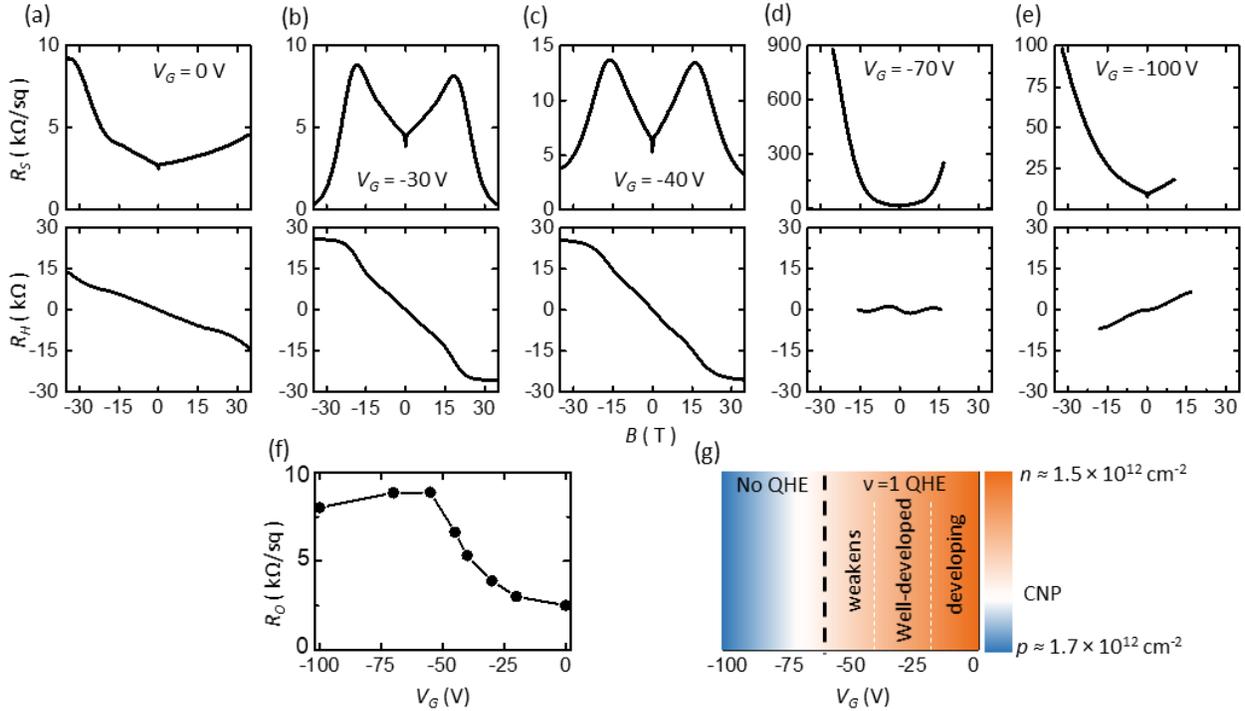

**Figure 2 | QHE as a function of magnetic field at several gate voltage values for a 10 QL film. (a)-(e)** sheet resistance ($R_S$, upper panel) and Hall resistance ($R_H$, lower panel) up to $|B| = 34.5$ T at $V_G = 0$, -30, -40, -70 and -100 V respectively. Change in the sign of Hall effect and corresponding maxima in **(f)** zero field sheet resistance $R_o$ at $V_G \approx -70$ V indicates that $p$-type carriers dominate the transport for $V_G < -70$ V. For $n$-type carriers, $\nu = 1$ QHE is clearly observed at high magnetic fields. **(g)** Evolution of QHE with carrier

density, where QHE disappears when carriers change from $n$- to $p$-type. $p$-type carrier density for $V_G = -100$ V was estimated from the average slope of Hall effect in (e).

Figures 2(a)-(e) show $R_S$ (upper panel) and $R_H$ (lower panel) of another identically prepared 10 QL film as a function of $B$ up to $\pm$ 34.5 T for various $V_G$ values at $T = 0.35$ K [15]. At low magnetic fields, magnitude of the negative slope of $R_H$ increases as $V_G$ changes from 0 to -40 V, indicating decreasing $n$-type carrier density from $1.5 \times 10^{12}$ cm$^{-2}$ to $5.9 \times 10^{11}$ cm$^{-2}$. At $V_G = -70$ V, $R_H$ fluctuates strongly around zero indicating mixed transport from electrons and holes and at $V_G = -100$ V the slope becomes positive albeit non-linear. The nonlinear Hall effect possibly indicates multi-carrier transport likely due to the effect of electron-hole puddles or due to residual $n$-type carriers on the top surface[5,21]. However, the overall positive slope of $R_H$ indicates that conduction is now dominated by $p$-type carriers. Figure 2(f) shows corresponding change in $R_o$, where it increases till $V_G = -55$ V, shows a maximum at $-70$ V $\leq V_G \leq -55$ V and then decreases below $V_G = -70$ V. This indicates that the film goes through CNP at $V_G \approx -70$ V.

At high magnetic fields, we observe increasingly developed dips in $R_S$ and plateaus at $h/e^2$ in $R_H$ for $-30$ V $\leq V_G \leq 0$ V indicating $v = 1$ QH state. Additionally, we observe developing plateau-like features at $R_H \approx h/3e^2$ in the same voltage range, which is consistent with top and bottom surfaces having similar carrier density in this gate voltage range [15]. At $V_G = -40$ V both the dip in $R_S$ and the plateau in $R_H$ are less well developed than at $V_G = -30$ V indicating that the $v = 1$ QH state weakens as the Fermi level is lowered toward CNP. For $V_G = -70$ V ($-100$ V), QH signature is completely gone and $R_S$ increases monotonically with $B$ and reaches ~878 k$\Omega$/sq (~90 k$\Omega$/sq) at $B = 25$ T (32 T), corresponding to ~10,000% (~1000%) of magnetoresistance as defined by MR% $= \frac{R_S(B) - R_S(0)}{R_S(0)} \times 100$ %. In Fig. 2(g), we summarize our observation, where $v = 1$ QHE emerges with decreasing $n$-type carrier density until $V_G \approx -30$ V, then diminishes while approaching CNP and gives way to a highly resistive state in the *p*-regime. The color plot in Fig. 2(g) was obtained by smoothly interpolating between the carrier densities measured at $V_G = 0$ V and at $V_G = -100$ V. We note that the *p*-type carrier density at $V_G = -100$ V in Fig. 2(g) was estimated by taking the average, rather than the low-field, slope of the Hall resistance.

In order to observe continuous evolution of transport with $V_G$, we have measured $R_S$ and $R_H$ as a function of $V_G$ at various magnetic fields $B$ from 0 to 44.5 T and various temperatures $T$ from 0.35 K to 9 K on another 10 QL thick film. Figures 3(a) and 3(b) show $R_S$ and $R_H$, respectively, at different magnetic fields and at $T = 0.35$ K [15]. For all magnetic fields we observe a peak in $R_S$ and a change in the sign of $R_H$ from negative to positive at $-30$ V $\lesssim V_G \lesssim -26$ V, indicating that *p*-type carriers dominate the transport below these gate voltages. For higher magnetic fields ($B > 23$ T), we observe a developing dip in $R_S$ and a plateau in $R_H \approx -25.8$ k$\Omega$ for $-20$ V $\lesssim V_G \lesssim -15$ V indicative of $\nu = 1$ QH state for *n*-type carriers. Near CNP and for *p*-type carriers, we observe neither the dips in $R_S$ nor the plateaus in $R_H$ in magnetic field up to 44.5 T. Inset of Fig. 3(a) shows the magnetic field dependence of $R_S$ at CNP ($R_{CNP}$) and at $V_G = -76$ V ($R_{-76V}$; where *p*-type carriers dominate) indicating that $R_S$ increases monotonically with $B$ for both CNP and *p*-type carriers, reaching as high as ~250 k$\Omega$ (44 k$\Omega$) and ~44 k$\Omega$ (18 k$\Omega$) at $B = 44.5$ T (11.5 T), respectively, which correspond to ~3,000% (600%) and 650% (100%) of magnetoresistance.

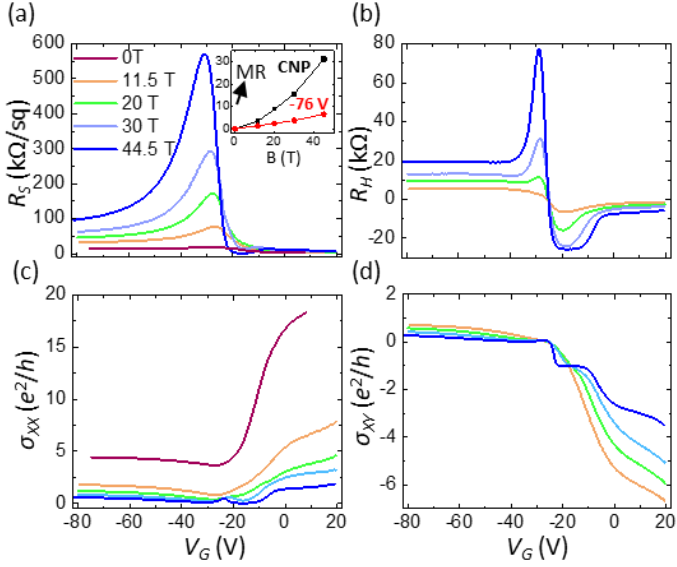

**Figure 3 | Gate voltage dependent transport properties of a 10 QL film at $T$= 0.35 K and $B$ field up to 44.5 T. (a)** sheet resistance ($R_S$) and **(b)** Hall resistance ($R_H$) at $T = 0.35$ K as a function of $V_G$ at several $B$ values from 0 to 44.5 T. Corresponding **(c)** sheet ($\sigma_{XX}$) and **(d)** Hall ($\sigma_{XY}$) conductance. $\nu = 1$ QHE is observed at $V_G \approx -15$ V to $-20$ V and non-saturating magnetoresistance (MR) $= \frac{R_S(B)-R_S(0)}{R_S(0)}$ with $B$ is observed for $V_G \lesssim -21$ V as plotted in inset of (a) for CNP and $V_G = -76$ V.

In order to get an additional perspective, we plot sheet conductance $\sigma_{XX} = R_S/(R_S^2 + R_H^2)$ and Hall conductance $\sigma_{XY} = R_H/(R_S^2 + R_H^2)$ in Fig. 3(c) and 3(d), respectively. Consistent with resistance plots, we observe a $\sigma_{XY}$ plateau at $\sim h/e^2$ and a minimum in $\sigma_{XX}$ at $-20$ V $\lesssim V_G \lesssim -15$ V indicative of $\nu = 1$ QHE. Apart from a plateau-like feature at $\sigma_{XY} \approx 0$ and corresponding minimum in $\sigma_{XX}$ for $-35$ V $\lesssim V_G \lesssim -30$ V, which can possibly indicate the $\nu = 0$ state, no features resembling QHE are observed in conductance plots for $V_G \lesssim -30$ V[4,5,9,22], implying that QHE is lost when $p$-type carriers dominate the transport. The lack of QHE on the $p$-side is consistent with both the recent transport result on compensation-doped $Bi_2Se_3$ films and the lack of Landau levels (LL) on the $p$-side of $Bi_2Se_3$ in scanning tunneling spectroscopy measurements[13,23,24]. It can be explained by the proximity of the Dirac point to the bulk valence band and the much broader surface band on the $p$-side of $Bi_2Se_3$[25,26]. This is in marked contrast with Sb-based TIs, which exhibit QHE and LLs for both $n$- and $p$-sides due to relatively symmetric surface bands with a well-exposed Dirac point[4,27,28,27].

Next, we discuss temperature dependence of $R_S$ versus $V_G$ at different magnetic fields in order to understand the behavior of $p$-type carries in $Bi_2Se_3$. As shown in Fig. 4(a), zero-field $R_S$ is < 7.5 kΩ/sq, which is much lower than the quantum resistance (25.8 kΩ), and increases only slightly (~4% at CNP) at $T = 0.35$ K compared to $T = 2$ K for all $V_G$. At higher temperatures ($T \approx$ 6 to 12 K), we observe similarly small variation in $R_S$ with temperature in a different but identically prepared 10 QL film [15]. In the inset of Fig. 4(a) we show $R_S$ versus $T$ at $V_G = 0$ V, where small upturn at low temperatures is observed, consistent with previous studies on TI films. In addition, we observe weak anti-localization at low fields for all $V_G$ values and in all samples (see Fig. 2 and 3). Both of these observations are consistent with gapless Dirac band transport in the presence of disorder[4,29].

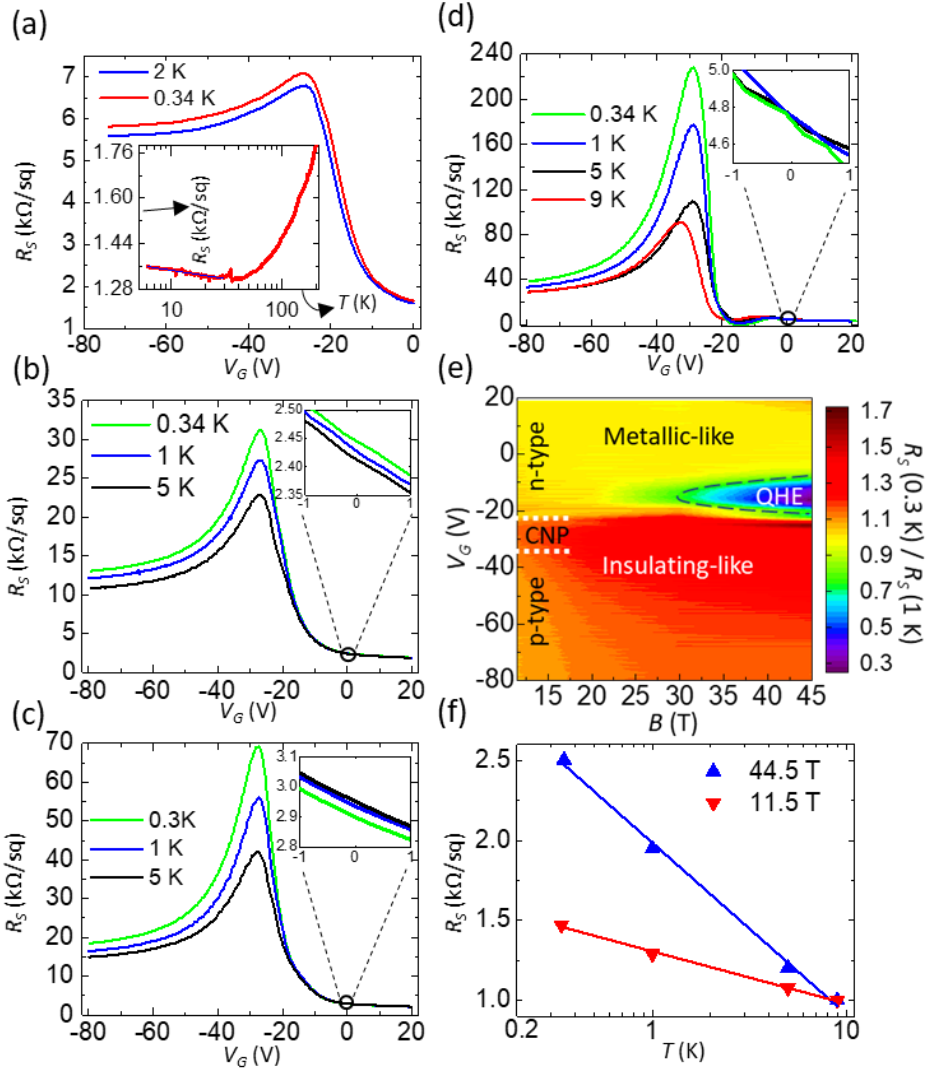

**Figure 4 | Temperature dependence of resistance at high magnetic field.** Sheet resistance ($R_S$) at $B =$ **(a)** 0 T **(b)** 11.5 T **(c)** 20 T and **(d)** 44.5 T as a function of $V_G$ at several different temperatures. Note that the film was not measured at 9 K in (c). Inset in (a) shows semi-log plot of $R_S$ versus $T$ at $V_G = 0$ V indicating a small upturn at low temperatures, which fits $\log(T)$ dependence as indicated by the blue line. Insets in (b-d) show behavior of $R_S$ around $V_G = 0$ V in greater detail. **(e)** 2D plot of ratio $R_S(0.34\ \text{K})/R_S(1\ \text{K})$ plotted as a function of $V_G$ and $B$. Metallic-like behavior is observed for $V_G > -21$ V while, for $V_G \lesssim -21$ V insulating-like behavior is observed. **(f)** Semi-log plot of temperature dependence of normalized $R_{CNP}$ at $B = 11.5$ T

and 44.5 T, where larger slope at 44.5 T indicates stronger insulating tendency: the solid lines are least-square fits using $R_{CNP} \sim \log(T)$.

In order to see how transport changes at higher fields, we plot $R_S$ versus $V_G$ at 11.5 T, 20 T and 44.5 T for $T = 0.35$ K – 9 K in Fig. 4 (b)-(d), respectively [15]. A slight shift in $V_G$ corresponding to CNP at $T = 9$ K compared to lower temperatures is observed, but does not affect our analysis. In Fig. 4 (b)-(d), $R_S(T_1) \approx R_S(T_2)$ for $V_G > -21$ V indicating metallic behavior, while for $V_G \lesssim -21$ V, $R_S(T_1) \gg R_S(T_2)$ suggesting an insulating behavior for these gate voltages for all three magnetic fields, where $T_1 < T_2$ are temperatures. Figure 4(e) summarizes this observation, where we show a 2D plot of the ratio of $R_S$ at $T = 0.3$ K to that at $T = 1$ K ($R_{0.3\ K}/R_{1\ K}$) as a function of $V_G$ and $B$. $n$-type carrier region shows metallic-like behavior (i.e. $R_{0.3\ K}/R_{1\ K} \approx 1$) along with QHE at high magnetic fields, while an insulator-like highly resistive state (i.e. $R_{0.3\ K}/R_{1\ K} > 1$) is observed near CNP and for $p$-type carrier region at high fields.

In order to further understand the nature of the highly-resistive state, we have plotted $R_{CNP}(T)/R_{CNP}(9\ K)$ as a function of temperature at $B = 11.5$ T and 44.5 T in Fig. 4(f). $R_{CNP}(T)/R_{CNP}(9\ K)$ increases logarithmically with decreasing temperature for both $B = 11.5$ T and 44.5 T, with stronger insulating behavior observed at higher field as indicated by greater slope of $R_{CNP}$ vs. $\log(T)$ for $B = 44.5$ T. For comparison, insets of Fig. 4 (b)-(d) show an enlarged view of $R_S$ versus $V_G$ at $V_G \approx 0$ V, where $R_S$ either decreases or does not increase significantly with decreasing temperature indicating a metallic behavior. Such an increasingly insulating behavior near CNP at higher magnetic fields indicates presence of a magnetic-field-induced insulator phase, whose origin remains unknown at present. Local and non-local measurements at lower temperatures and higher magnetic fields could elucidate the nature of this ground state, which we leave for future work.

In conclusion, we have studied gate-dependent QHE on low-carrier density $Bi_2Se_3$ thin films for the first time by employing a novel buffer layer growth method on gate-amenable $SrTiO_3$ substrates. At low fields we observe ambipolar transport for thinner films, and at high fields we observe $v = 1$ QHE for *n*-type carriers. On the other hand, for CNP and *p*-type carriers we observe non-saturating magnetoresistance up to $B = 44.5$ T, whose temperature dependence point to the existence of a magnetic-field-induced insulating state. Further experimental and theoretical efforts are necessary to clarify its origin.

**Acknowledgements:**

This work is supported by the Gordon and Betty Moore Foundation's EPiQS Initiative (GBMF4418) and National Science Foundation (NSF) grant EFMA-1542798. A portion of this work was performed at the National High Magnetic Field Laboratory which is supported by NSF Cooperative Agreement No. DMR-1644779 and the State of Florida.

**Author contributions:**

N.K. and S.O. conceived the experiment. N.K., M.S. and J.M. synthesized the samples and performed transport measurements. N.K. and S.O. wrote the manuscript with inputs from all the authors. All authors contributed to the scientific analysis and manuscript revisions.